\documentclass[twocolumn,showpacs,prl,floatfix,amssymb]{revtex4}

\usepackage{graphicx}

\begin{document}

\newlength{\plotwidth}          
\setlength{\plotwidth}{8.3cm}   

\title{Phase measurements using Two-Channel Fano Interference \\ in a Semiconductor Quantum Dot}

\author{C.~F\"uhner}
\email{fuehner@nano.uni-hannover.de}
\author{U.~F.~Keyser}
\author{R.~J.~Haug}
\affiliation{Institut f\"ur Festk\"orperphysik, Universit\"at
Hannover, Appelstr. 2, 30167 Hannover, Germany}

\author{D.~Reuter}
\author{A.~D.~Wieck}%
\affiliation{Lehrstuhl f\"u{}r Angewandte Festk\"o{}rperphysik,
Ruhr-Universit\"at Bochum, 44780 Bochum, Germany}

\date{\today}

\begin{abstract}
We investigate a lateral semiconductor quantum dot with a large
number of electrons in the semi-open Fano regime.  In transport
measurements we observe three stable series of Fano resonances
with similar lineshapes.  We present a simple model explaining the
temperature and $V_{SD}$ dependence of the resonances. The Fano
regime allows to investigate phase and coherence of the electronic
wave function and astonishingly, we find no signs of decoherence
in our system.
\end{abstract}

\pacs{72.15.Qm, 73.21.La, 73.23.Hk, 73.40.Gk}

\maketitle


Transport through semiconductor quantum dots \cite{Kouwenhoven-97}
has been studied in a number of different regimes. Each regime is
characterized by specific lineshapes of the resonances in linear
conductance measurements. In the Coulomb regime at weak tunnel
coupling between dot and leads basically Breit-Wigner type
resonances are observed \cite{Beenakker-91}. At stronger coupling
an increased valley conductance indicates the Kondo regime
\cite{Kondo-Alle}. Experiments with transparent barriers in the
open regime are interpreted in terms of ballistic transport with
electrons geometrically reflected within the dot \cite{Bird-03}.
Here, we explore an intermediate regime, the semi-open Fano regime
where a ballistic transport channel coexists with a resonant
transport channel from Coulomb blockade.

\begin{figure} 
\begin{center}
  \resizebox{\plotwidth}{!}{\includegraphics{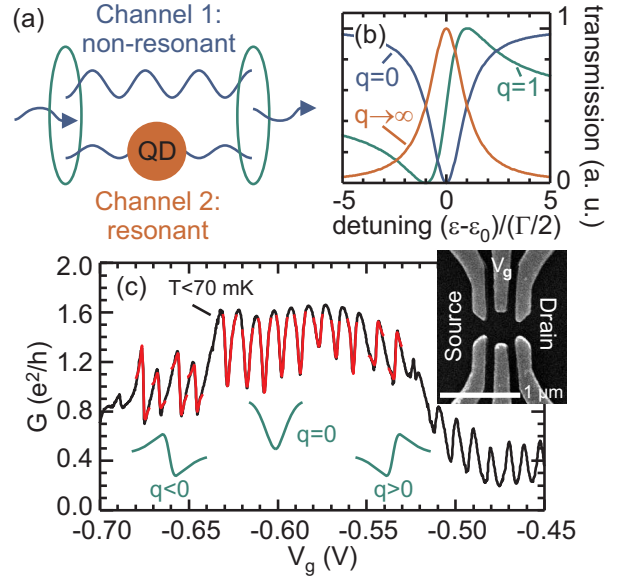}}
\end{center}
\caption{(a) Model system for Fano resonances.  An incident wave
function interferes after transmission through two channels, a
resonant and a non-resonant one. Here, the resonant channel is
formed by a quantum dot (QD). (b) Fano transmission lineshapes
parameterized by $q$. (c) Linear conductance $G$ of our quantum
dot versus plunger gate voltage $V_g$ in the semi-open Fano regime
exhibiting three regimes with $q<0$, $q=0$, and $q>0$. Fits to the
individual resonances are shown in red. Inset: SEM picture of the
sample.} \label{fig1}
\end{figure}

Fano resonances are a very general characteristic of systems where
two transmission channels interfere, a resonant and a non-resonant
one (Fig.~\ref{fig1}(a)).  Depending on the relative phase and
amplitude of the two channels they exhibit a wide range of
lineshapes parameterized by the asymmetry parameter $q$
(Fig.~\ref{fig1}(b)).  The analytical expression for Fano
resonances is \cite{Fano-61}:
\begin{equation}
  f_{Fano}(\tilde\varepsilon) = \frac{(\tilde\varepsilon + q)^2}{\tilde\varepsilon^2 + 1}
  \label{eqFano}
\end{equation}
with $\tilde \varepsilon = (\varepsilon - \varepsilon_0) / (\hbar
\Gamma/2)$ the detuning from the resonance at energy
$\varepsilon_0$ and normalized to the width $\Gamma$ of the
resonance. As shown in Fig.~\ref{fig1}(b), we get Breit-Wigner
type peaks for $q\to\infty$, dips for $q=0$, and asymmetric
lineshapes in between as shown for $q=1$. Negative
parameters $q$ lead to lineshapes mirrored at the ordinate, i.~e.\
$\tilde\varepsilon \to -\tilde\varepsilon$.

Fano theory was initially developed in the context of
electron-atom scattering \cite{Fano-61} but later applied to a
large number of other experiments, ranging from electron-neutron
scattering \cite{Neutron-Alle} to atomic photoionization
\cite{Fano-65}. More recent observations of Fano resonances in
electronic transport experiments include crossed multiwall carbon
nanotubes \cite{Kim-03}, scanning tunneling microscopy of single
adatoms on a metallic surface \cite{STM-Alle} and transport
studies of man-made semiconductor nanostructures \cite{Nockel-94}.
Kobayashi and coworkers observed Fano resonances in an
Aharonov-Bohm ring with a quantum dot embedded in one arm
\cite{Kobayashi-02}. The resonant channel in their system stems
from Coulomb blockade in the quantum dot, and the non-resonant
channel originates from the continuous spectrum of the other arm.

In contrast, in pure quantum dot experiments in the semi-open
regime the dot itself can be the interferometer. There is no
spatial separation between the two transport channels. The
resonant transmission mechanism is the same as known from Coulomb
blockade. An additional non-resonant transmission channel comes
into existence when the time needed to traverse the dot via a
direct trajectory, $\Delta t \sim l/v_F$, is small enough ($l$ is
the length of the path and $v_F$ the Fermi velocity). Then by the
time-energy uncertainty principle with the Coulomb charging energy
$U$, $\Delta t \cdot U \ge h$, the transmission via the direct
path is no longer forbidden by Coulomb blockade. So far, there
have been only few experimental studies on quantum dots in the
semi-open regime \cite{Gores-00,Zacharia-01}. However, these
experiments have generated great theoretical interest
\cite{Racec-01,Xiong-02,Bulka-01,Hofstetter-01}.  They are
sensitive tools to gain information on the phase of transmitted
electrons inaccessible in other regimes \cite{Clerk-01,Clerk-01b}.
It was even suggested to use Fano resonances as a probe of phase
coherence in quantum dots.


In this paper, we study a quantum dot in the semi-open Fano regime
in low-temperature transport measurements.  We extract
quantitative information of the two interfering channels including
phase.  Decoherence is studied by application of finite
temperature and source-drain voltage.



We fabricated our sample from a high-mobility GaAs/AlGaAs
two-dimensional electron system (2DES) with an electron density $n
= 3.7 \cdot 10^{15}$~m$^{-2}$ and a mobility $\mu = 130$~m$^2$/Vs.
Cr/Au split-gates were patterned by electron beam lithography
(Fig.~\ref{fig1}(c), inset). We apply negative gate voltages to
form a quantum dot tunnel coupled to 2DES leads. The plunger gate
voltage $V_g$ controls the dot energy levels and -- due to spatial
proximity to the tunnel barriers -- to some extent also influences
the coupling of the dot to the leads. We use a standard lock-in
technique for differential conductance measurements in a
$^3$He-$^4$He dilution refrigerator with an electronic base
temperature less than 70~mK.


We can tune our quantum dot from the closed Coulomb regime to the
semi-open Fano regime. Linear conductance measurements in the Fano
regime are shown in Fig.\ \ref{fig1}(c). We clearly observe Fano
anti-resonances and asymmetric resonances instead of Breit-Wigner
type peaks known from Coulomb blockade. We find three groups of
resonances with similar lineshapes corresponding to asymmetry
parameters $q<0$, $q=0$, and $q>0$ in the gate voltage range from
$V_g=-0.68$~V to $V_g=-0.52$~V. This is different from the
continuous evolution of lineshapes in multi-level interference
experiments \cite{Lindemann-02}. For $V_g>-0.52$~V the
Breit-Wigner lineshape is restored. In the Fano regions, the
background conductance is high ($\sim 1-2$~$e^2/h$) similar to the
open regime.

\begin{figure} 
\begin{center}
  \resizebox{\plotwidth}{!}{\includegraphics{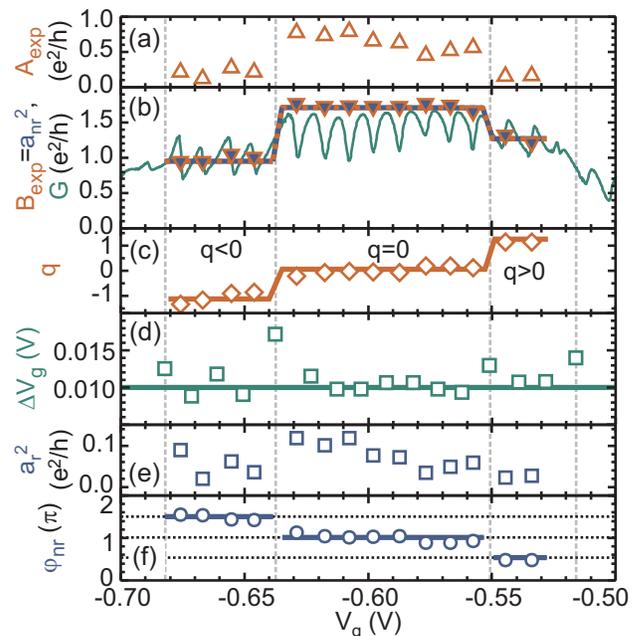}}
\end{center}
\caption{(a)--(c) Parameters $A_{exp}$, $B_{exp}$, and $q$ for
fits of individual Fano resonances from Fig. \ref{fig1}(c) to eq.\
\ref{eqFanoFit} (red).  The original linear conductance data is
also shown in (b) for reference (green). The vertical lines
separate the three groups of similar Fano resonances, the others
are guides to the eye. (d) The resonance-to-resonance spacing
$\Delta V_g$ (squares) is nearly constant as highlighted by the
red line. Exceptions coincide with changes in $q$ at the vertical
grey lines.  (b),(d),(e) show the extracted parameters $a_{nr}^2$,
$a_r^2$, and $\varphi_{nr}$ of the two-channel model (blue).}
\label{fig2}
\end{figure}

To fit the measured Fano resonances (red curves in Fig.\
\ref{fig1}(c)), we use eq.\ \ref{eqFano} with an additional
prefactor and offset,
\begin{equation} \label{eqFanoFit}
  f_{exp}(V_g) = A_{exp} \cdot [ f_{Fano}(\tilde\varepsilon(V_g)) - 1 ]
   + B_{exp}.
\end{equation}
$A_{exp}$ and $B_{exp}$ account for a reduced transmission
amplitude of the two interfering channels, e.~g.\ due to
backscattering in one channel, and for an incoherent contribution
to the total conductance.  The fit parameters $A_{exp}$, $B_{exp}$
and $q$ are plotted in Fig.\ \ref{fig2}(a)--(c).  The evolution of
the parameters exhibits clear steps at the positions marked by the
vertical grey lines. The parameters are nearly constant within one
group of resonances but change in between.

This switching is further investigated in Fig.\ \ref{fig2}(d)
where we have analyzed the spacing $\Delta V_g$ of the resonances
in gate voltage.  The spacing is nearly constant within any group
of resonances but shows increased values at the switching points
of the fit parameters. In the two-channel model (Fig.\
\ref{fig1}(a)), the positions of the Fano resonances are
determined by the $V_g$ positions of the Coulomb resonances in the
resonant channel. In a large dot like ours, the spacing of Coulomb
peaks is dominated by the charging energy $U$ which is large
compared to the single-particle energy level spacing $\Delta E_i$,
$U > \Delta E_i$. Clear deviations from a constant peak spacing
are thus attributed to greater redistributions of the charge on
the dot. These redistributions influence the non-resonant
transmission channel which also traverses the dot. Consequently,
the interference of both channels and hence the Fano lineshape is
altered.  Within the groups of constantly spaced and similarly
shaped Fano resonances, the transmission of the non-resonant
channel is constant.


For a quantitative understanding of the Fano resonances, we
investigate the model of two interfering channels in more detail:
In the non-resonant conductance channel, the electronic wave
function evolves like a plane wave. After traversing the channel
its amplitude is $a_{nr}$ and its phase is $\varphi_{nr}$.
Compared to the width of a resonance, the energy dependence of
$a_{nr}$ and $\varphi_{nr}$ is negligible. Thus the complex
transmission of the channel is:
\begin{equation} \label{eqtnr}
  t_{nr} = a_{nr} \cdot \exp(i \varphi_{nr}).
\end{equation}
$t_{nr}$ is modified in charge reconfigurations on the dot but
otherwise constant as discussed above.

To model the complex transmission of the resonant channel, we
consider a Coulomb resonance at zero temperature ($T=0$).
According to experiments by Schuster {\em et al.}
\cite{Schuster-97}
we use the Breit-Wigner formula with a normalized detuning
$\tilde\varepsilon$ from resonance and with an amplitude $a_r$:
\begin{equation} \label{eqtr}
  t_{r}(\tilde\varepsilon) =
    a_r \cdot \frac{i}{\tilde\varepsilon + i}
\end{equation}
The transmission probability
$T_r(\tilde\varepsilon)=|t_r(\tilde\varepsilon)|^2=a_r^2/(\tilde\varepsilon^2+1)$
(i.~e.\ the conductance) of this channel yields the well-known
Breit-Wigner peaks. The total transmission amplitude
\begin{equation} \label{eqTtot}
  T_{tot}(\tilde\varepsilon) = | t_{nr} + t_{r}(\tilde\varepsilon) |^2.
\end{equation}
describes the conductance of the complete Fano system from Fig.\
\ref{fig1}(a) including interference. The comparison of equations
\ref{eqFanoFit} and \ref{eqTtot} allows to map the
phenomenological constants $(A_{exp}, B_{exp}, q)$ determined from
the fit to the more physical parameters $(a_{nr},a_r,\varphi_nr)$.
It is easily shown that $a_{nr}^2 = B_{exp}$. The other
expressions are more complex due to trigonometric functions.

The transformation of the experimental and phenomenological data
from Fig.\ \ref{fig2} is shown in the same figure in blue. The
non-resonant transmission amplitude $a_{nr}$ and phase
$\varphi_{nr}$ as functions of gate voltage reflect the switching
in the non-resonant channel.  Phase differences of
$\Delta\varphi_{nr}=\pi/2$ correspond to a length difference of at
least $\Delta\varphi_{nr} / k_F = 10$~nm ($k_F$ is the Fermi
wavevector) between non-resonant channels for different charge
configurations. The transmission amplitude $a_{nr}$ varies between
$1.0$ and $1.7$~$e^2/h$. It thus can be explained by a
one-dimensional spin-degenerate channel with its conductance of
2~$e^2/h$ slightly reduced due to backscattering. The
backscattering is modified at the steps. In contrast the resonant
channel's amplitude $a_r^2$ exhibits an unsystematic modulation
which is well known from Coulomb blockade and is caused by the
varying overlaps of wave functions for different numbers of
electrons.


\begin{figure} 
\begin{center}
  \resizebox{\plotwidth}{!}{\includegraphics{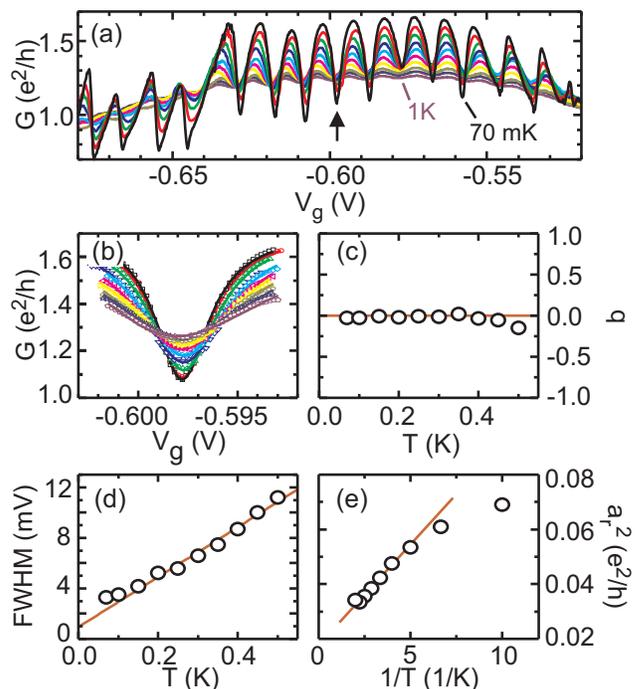}}
\end{center}
\caption{Fano resonances in the linear conductance for
temperatures between $T=70$~mK (black) and $T=1$~K (violet). (b)
Single Fano anti-resonance marked in (a) for temperatures
$T=70$~mK -- 1~K fitted by eq. \ref{eqFanoFit}. The temperature
dependence of this resonance is further investigated under the
following aspects: (c) Asymmetry parameter $q$, (d) full width at
half minimum (FWHM), and (e) transmission $a_r^2$ of the resonant
channel. The temperature scale in (e) is reciprocal.} \label{fig3}
\end{figure}

We have studied how temperature affects Fano resonances. In
Fig.~\ref{fig3}(a) the resonances are shown for temperatures from
$T=70$~mK to $T=1$~K. Although the resonance amplitude decreases
with increasing temperature, the lineshape described by $q$
clearly remains unaffected.  One of the resonances from
Fig.~\ref{fig3}(a) is analyzed more closely in
Fig.~\ref{fig3}(b)--(e) in much the same way as in
Fig.~\ref{fig2}. We have fit the resonance with
eq.~\ref{eqFanoFit} and found $q$ to be independent of temperature
up to 0.4~K (Fig.~\ref{fig3}(c)). This is understood in the
two-channel model where the non-resonant channel is temperature
independent. The resonant channel has all properties of Coulomb
blockade including temperature dependence which is due to the
Fermi-Dirac broadening of the electrons in the emitter
\cite{Beenakker-91}. However, the relative transmission amplitudes
and, most important, the phase evolution of both transmission
channels remain unaffected. This is why the shape of the
resonances and $q$ are constant. Only width and amplitude are
influenced by the temperature as determined by the emitter.

To further verify the two-channel model we analyze the width
$\Gamma$ of the Fano resonances which is contained in $\tilde
\varepsilon = (\varepsilon - \varepsilon_0) / (\hbar \Gamma/2)$ in
eq.~\ref{eqFanoFit}. Fig.~\ref{fig3}(d) shows that $\Gamma$
increases linear with temperature. As the non-resonant channel is
$T$-independent, the broadening of the Fano interference directly
reflects a broadening of the Coulomb channel. In Coulomb blockade,
the slope of $\Gamma(T)$ is $3.5 k_B T$ due to Fermi broadening of
the emitter \cite{Beenakker-91}. Thus from the slope in
Fig.~\ref{fig3}(d) we can extract a conversion factor $\alpha =
0.015$ between plunger gate voltage and chemical potential, $\mu =
\alpha e V_g$. This has to be compared to $\alpha = 0.045$ from
the Coulomb regime at $V_g=-1.15$~V. The difference is attributed
to the increased source and drain capacitances at the larger
tunnel coupling in the Fano regime.

In Fig.~\ref{fig3}(e), the amplitude of the resonant channel
$a_{r}^2$ is analyzed as a function of temperature. On a
reciprocal scale, the amplitude of the resonant channel decreases
linearly with increasing temperature as known from Coulomb
blockade \cite{Beenakker-91}. Deviations are from thermal
decoupling of the electron system from the bath for low
temperatures. Please note that as we examine Fano dips originating
from destructive interference, a decrease in amplitude in the
resonant channel results in an increased total conductance as
observed in Fig.~\ref{fig3}(b). This mechanism is quite different
from the logarithmic temperature dependence of Kondo valleys
\cite{Kondo-Alle} which might look similar.

At temperatures higher than 0.4~K the overlap of the Fano
resonances becomes so strong that our analysis becomes unreliable.
It is then impossible to determine the width or amplitude of
individual peaks.

The temperature dependence of Fano resonances thus is completely
understood in terms of the two-channel model. It results from the
Fermi-Dirac broadening of the emitter and its influence on the
transmission amplitude and width of the resonant channel.
Apparently Fermi-Dirac broadening of the emitter does not
influence the non-resonant channel. Our results show that
decoherence between both channels can be neglected in our Fano
system in the investigated temperature range. Decoherence between
both channels would result in a breakdown of the interference. The
resonances would then consist of the well-known resonant Coulomb
peaks plus an offset from the non-resonant background. Kobayashi
and coworkers observed an evolution of Fano resonances to Coulomb
peaks plus background at $T \sim 200$~mK in their Aharonov-Bohm
ring with an embedded quantum dot \cite{Kobayashi-02}. We
attribute the strong influence of decoherence in their Fano system
to the large size of their system (circumference $\sim 4$~$\mu$m).


\begin{figure} 
\begin{center}
  \resizebox{\plotwidth}{!}{\includegraphics{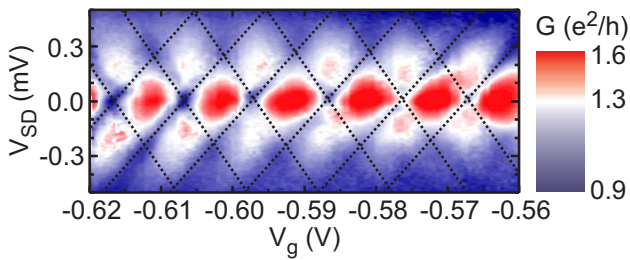}}
\end{center}
\caption{Nonlinear differential conductance as a function of
plunger gate voltage $V_g$ and bias voltage $V_{SD}$ in the $q=0$
regime of Fano anti-resonances. The diamonds of low conductance
known from Coulomb blockade have been replaced by anti-diamonds of
high conductance. These are highlighted by black dotted lines.}
\label{fig4}
\end{figure}

Coherence can also be destroyed by finite source-drain voltages.
For finite bias voltages quantum dots show so-called Coulomb
diamonds in non-linear differential conductance measurements
versus source-drain voltage $V_{SD}$ and gate voltage $V_g$. These
are diamond-shaped areas of low conductance. In our Fano quantum
dot, however, we observe clear anti-diamonds of {\em high}
conductance for the $q=0$ regime (Fig.~\ref{fig4}). These are
again understood in the two-channel model.  The resonant channel
based on Coulomb blockade determines the resonance positions.  The
total conductance, however, is determined by the interference of
both conductance channels, which leads to anti-diamonds in the
$q=0$ anti-resonance regime. Thus the appearance of anti-diamonds
confirms our interpretation of the Fano regime.  Non-linear
differential conductance measurements also address decoherence.
From Fig.~\ref{fig4} it is clear that at least for $V_{SD} = -0.3$
-- 0.3~mV the interference is preseved. The decoherence introduced
by the finite source-drain voltage $V_{SD}$ is not sufficient to
destroy the interference. Finally, Fig.~\ref{fig4} demonstrates
the low charging energy in the Fano regime of $U \sim 0.3$~eV.
Insertion into the Heisenberg uncertainty relation yields roughly
$\Delta t \cdot U \sim 0.1 h$. This is consistent with the
existence of a direct, virtual, and non-resonant transmission path
not respecting Coulomb blockade.



In conclusion, we explored the semi-open Fano regime of quantum
dots.  We found stable series of resonances with similar
lineshapes.  We have interpreted our measurements in terms of two
interfering transmission channel and extracted phase information
for the non-resonant channel.  The sudden changes in lineshape
were interpreted as charge reconfigurations on the dot influencing
the direct non-resonant trajectory.  We found our Fano
interferometer robust with respect to decoherence in temperature
and source-drain voltage dependent measurements.


We thank F.~Hohls and U.~Zeitler for help with the measurement
setup and acknowledge financial support by BMBF.






\newpage




\end{document}